\documentstyle[12pt,epsfig]{article}

\begin{document}
  \newcommand{\ccaption}[2]{
    \begin{center}
    \parbox{0.85\textwidth}{
      \caption[#1]{\small{{#2}}}
      }
    \end{center}
    }

\vspace{1truecm}
\begin{center}
{\large { \bf
ASYMMETRY IN CHARMED PARTICLES PRODUCTION IN $\Sigma^{-}$ BEAM }}
\end{center}

\vspace*{0.5truecm}

\begin{center}
A.K.~Likhoded$^1$~and~S.R.~Slabospitsky \\
State Research Center \\
Institute for High Energy Physics, \\
Protvino, Moscow Region 142284, \\ RUSSIA
\end{center}

\vspace*{1.5truecm}

\begin{center}
Abstract
\end{center}
We present the calculation of the inclusive $x_F$-distributions of charmed 
hadrons, produced in high-energy $\Sigma^-$-beam. The calculation is based on 
the modified mechanism of charmed quarks fragmentation as well as on the 
mechanism of $c$-quark recombination with the valence quarks from initial 
hadrons. We predict the additional asymmetry in the production of charmed
hadrons  due to the different distributions of the valence
$s$ and $d$ quarks in $\Sigma^-$-beam. 

\vfill

\rule{3cm}{0.5pt}

$^1$E--mail:~LIKHODED$@$mx.ihep.su,

\newpage

\section* {\bf Introduction}
 
In the hadronic production of particles with open charm an interesting
situation, connected with the interaction of charmed quarks with the hadronic
remnant, takes place~(see~\cite{expall,Adamovich:1999mu, selex} and 
references therein). 

Let us remind that such a problem does not arise in the case of $e^+ e^-$
annihilation, where a heavy quark is hadronized due to its own radiation and
all the description is reduced to the fragmentation function: 
\begin{equation} \label{ee}
\frac{1}{\sigma_{c \bar c}} \frac{d \, \sigma}{d \, z} = D(z, \, Q^2)
\nonumber 
\end{equation}
here $z$ is the part of the $c$-quark momentum, taken by charmed hadron and 
$Q^2$ is the square of total energy in the $e^+ e^-$ annihilation.

It is well known that a simple transition of factorization approach in the form
of (\ref{ee}) to the charmed particle production in hadronic interaction 
 leads to the valuable deviation from
experimental data. Indeed, there is a substantial difference in the yield of
different charmed hadrons in the fragmentation region $(|x| \to 1,
\quad x \equiv 2 p_H /\sqrt{s})$ of initial hadrons. 

This charmed particle yields asymmetry $A$ (or {\bf ``leading particle 
effect''}) is defined as follows:
\begin{eqnarray}
A = \frac{ 
 \frac{d \sigma}{dx}(leading) \; - \; \frac{d \sigma}{dx}(nonleading) }
 {\frac{d \sigma}{dx}(leading) \; + \; \frac{d \sigma}{dx}(nonleading) }. 
\label{asym}
\end{eqnarray}
Here symbol 'leading' or 'non-leading' refers to charmed hadron with or 
without light quarks identical to valence quarks from initial hadrons. For
example, in $\pi^-({\bf \bar u d})$-beam the mesons $D^0(c \bar {\bf u})$
and  $D^-(\bar c  {\bf d})$ are leading particles, while 
$\bar D^0(\bar c {\bf u})$ and  $D^+(c \bar {\bf d})$ should be considered
 as non-leading ones.
 
There are many theoretical articles devoted to description of a such 
phenomenon~\cite{models}. In our previous~\cite{ls1, Likhoded} publications we
 have taken
into account the interaction of $c$--quarks with quarks of the initial 
hadrons and
obtained a good agreement with the data in the spectra description at large $x$
and the $x$--asymmetry dependence in $\pi^- N~$ and 
$\Sigma^- N$-interactions.

The recent data on the charm production in the 
 $\Sigma^-$-beams~\cite{Adamovich:1999mu, selex} 
are of special interest due to the fact that the beam hadron (here $\Sigma$)
has $s$-quark and as consequence the distribution functions of the
valence quarks in the $\Sigma^-$-baryon differ from that of 
proton~\cite{Kiselev:1986at}. 
It has to lead to the set of observable effects and, in particular, to the
another asymmetry in the yield of the charmed hadrons that in 
the case of proton-proton collisions.

Preliminary data presented by the SELEX~Collaboration~\cite{selex} on the 
charm yield in the $\Sigma^- p$ and $\pi^- p$
collisions  show however some disagreement with other experiments
~\cite{Adamovich:1999mu} and general theoretical assumptions, in particular, 
in the expected charmed particle and anti-particle yields in the central region.

In this article we consider very briefly this problem from general theoretical
point of view (Section~1). In the Section~2 we give a short description 
of the fragmentation mechanism. Our approach for calculation of 
$x$-distributions of charmed-particle 
production in recombination mechanism is given in the Section~3. 
We  present the predictions for differential distributions
 of charmed hadrons produced in $\Sigma^- N$-interactions in the Section~4.
 A short summary of the results is given in the Conclusion.

\section {\bf Asymptotic behavior in the central region} 

Before discussing in detail of the spectra behavior and the charge asymmetry
in the hadron yield let us consider general behavior of charmed particle
spectra.

According to the generalized optical theorem ~\cite{Mueller} an inclusive
spectrum is connected with the discontinuity of the sixth particle amplitude by
the following relation~(see~Fig.~1.): 
\begin{equation}
E\frac{d^3 \sigma}{d^3 p} = \frac{1}{s} \, {\rm disc} M_{3\to3}, \label{reg1}
\end{equation}
where $s$ is the total energy squared.

Let us introduce two invariants (see~Fig.1)
\begin{eqnarray*}
 s_1 = (p_a + p_c)^2, \quad s_2 = (p_b + p_c)^2.
\end{eqnarray*}
At the high energies the asymptotic behavior of the $M_{3 \to 3}$ amplitude
as a function of these invariants $s_{1,2}$ is determined by leading
Regge trajectories, namely, the Pomeron~(${\cal P}$) and secondary 
trajectories~(${\cal R}_i$), related to $\rho$, $\omega$, $f$, $A_2$ mesons.
In so doing, the ${\cal PP}$, ${\cal RP}$, and 
${\cal RR}$ contributions become important.

In the central region $(x \sim 0)$, where the  kinematic invariants $s_1$
 and $s_2$ are large, one has
\begin{eqnarray}
 s_1 \approx \sqrt{s} \, m_{\bot} \, e^{-y}, \quad
 s_2 \approx \sqrt{s} \, m_{\bot} \, e^{y}, \label{s1s2} 
\end{eqnarray}
where $y$ is the rapidity and $m_{\bot} = \sqrt{m^2 + p_{\top}^2}$.

\noindent
Double Regge representation is a good approximation for the $M_{3 \to 3}$ 
amplitude (see~Fig.1) in this kinematic region. So, one has
\begin{equation}
E\frac{d^3 \sigma}{d^3 p} \approx \frac{1}{s} \, \sum_{i,j} 
 \tilde f_{ij} s_1^{\alpha_i}
 s_2^{\alpha_j} \label{reg2}
\end{equation}
where $\alpha_i$ and $\alpha_j$ are the intercepts of the 
leading Regge-trajectories~\cite{Mueller, collin}.  For Pomeron trajectory
one has  $\alpha_P \approx 1$, while the $f, \rho, \omega, A_2$ 
trajectories have the intercept  $\alpha_R \approx 1/2$.

The $\tilde f_{ij}(m_{\bot})$  are unknown 
functions of $m_{\bot}$, but it seems that these 
functions are universal and do not depend on the type of produced 
particles~\cite{Mueller, Kobrinsky:1975sy}. These functions can be 
determined by fitting to experimental data.
All the dependence on
quantum numbers of initial and final particles is determined by 
 the coupling  constants of the secondary Reggeons ${\cal R}$, included to 
the definition of the  $\tilde f_{ij}(m_{\bot})$ functions.

After substituting $s_{1,2}$ from (\ref{s1s2}) the expression (\ref{reg2}) 
takes the form:
\begin{equation}
E\frac{d^3 \sigma}{d^3 p} = \tilde f_{\cal PP}(m_{\bot}) - \sum_{i,j} 
 \frac{ \tilde f_{ij}(m_{\bot}) }{s^{1 - \frac{\alpha_i + \alpha_j}{2}}} 
 e^{(\alpha_j - \alpha_i)y} \label{reg3}
\end{equation}
This equation provides a good description of the transition to the asymptotic
 regime in the ``central'' region for all variety of the yields of  
$\pi$- and $K$-mesons~\cite{Kobrinsky:1975sy}.

For the $D$- and $B$-meson cases the expression~(\ref{reg3}) simplifies 
due to the fact that the contribution of the trajectories, connected with 
a charmed $c$-quark in $D$-meson ($J/\psi, \chi_c$ trajectories) or beauty 
$b$-quark in $B$-meson ($\Upsilon ,\; \chi_b$ trajectories), is strongly
suppressed by the Zweig rule for the coupling with initial mesons or nucleons. 
Therefore, in the sum of Eq.~(\ref{reg3}) there are no contributions with two
secondary Reggeons $({\cal RR}$-contributions) in the upper and bottom 
 ``shoulder'' of the diagram in Fig.~1.
So, one gets more simple expression:
\begin{equation}
E\frac{d^3 \sigma(D)}{d^3 p} \approx \tilde f_{\cal PP}(m_{\bot}) - 
\frac{1}{^4\sqrt{s}} \{\tilde f_{\cal RP}(m_{\bot}) e^{y/2} +
 \tilde f_{\cal PR}(m_{\bot}) e^{-y/2} \}. \label{reg4}
\end{equation}

Note, that the  contribution of the first ${\cal PP}$-term in the
equation~(\ref{reg4}) is the same for the particle and
 anti-particle
and  is proportional to the total cross-section of the interaction of
initial hadrons. Thus, the normalized cross-section 
\[ \frac{1}{\sigma_{tot}^{hh}} E\frac{d^3 \sigma(hh\,\to\,D\,X)}{d^3 p}
\]
at high energy limit has one universal limiting value  independently on the 
type of colliding hadrons.

For the case of the  $D_s$-meson production in the  $\Sigma$ or $K$-beams
the leading intercept  is related to $\phi$-meson, 
$\alpha_j = \alpha_{\phi} \approx 0$ and there is no
contribution from the third ${\cal PR}$-term in Eq.~(\ref{reg4}). The 
$\phi$-trajectory can be connected with $\Sigma$-particle only, because there 
are no valence strange quarks in the proton. As a result, one has only two 
terms: 
\begin{equation}
\frac{1}{\sigma_{tot}}E\frac{d^3 \sigma(D_s)}{d^3 p} \approx 
\tilde f_{\cal PP}(m_{\bot}) 
- \frac{1}{\sqrt{s}} \tilde f_{\cal RP}(m_{\bot}) e^{y} \label{reg5}
\end{equation}

The contribution of secondary trajectories determines the difference between
the yields of the particle and anti-particle in the 
central region. For instance, for $p\,p$ and $p \bar p$-collisions one has
\begin{displaymath}
\Delta_{p\,p} \sim \frac{a}{s^{1/4}} ch\frac{y}{2}, \quad
\Delta_{p\, \bar p} \sim \frac{b}{\sqrt{s}} sh\frac{y}{2}\;,
\end{displaymath}
where  the $a$ and $b$ coefficients depend on both combinations of secondary 
Reggeon  couplings and particle type of $D$- or $B$-meson. 

In the most experiments on fixed target the point $y=0$~$(x=0)$, as a rule, 
is out of reach
due to the experiment conditions, and the  measured spectrum starts
from $x \sim 0.1$. With a simple substitution 
$y \simeq \ln(x \frac{\sqrt{s}}{m_{\bot}})$
for $D$-mesons the equation~(\ref{reg4}) takes the form
\begin{equation}
\frac{1}{\sigma_{tot}}E\frac{d^3 \sigma(D)}{d^3 p} \simeq \tilde 
f_{\cal PP}(m_{\bot}) -
\{ \tilde f_{\cal RP}(m_{\bot})\sqrt{m_{\bot}}\sqrt{x} + 
\tilde f_{\cal PR}(m_{\bot}) \frac{ \sqrt{m_{\bot}}}{\sqrt{sx}} \}\;.
\end{equation}
For $D_s$-meson production in the $\Sigma$ or $K$ beams one has: 
\begin{equation}
\frac{1}{\sigma_{tot}}E\frac{d^3 \sigma(D_s)}{d^3 p} \simeq 
 \tilde f_{\cal PP}(m_{\bot}) - \tilde f_{\cal RP}(m_{\bot}) m_{\bot} x
\end{equation}
The asymmetry in the particle yield, which depends on the quantum numbers of 
initial hadrons and observed charmed particles, is determined by the second 
term in these expressions, which is different  for the
$D$ and $D_s$ cases. One can see that
the transition to the asymptotics at fixed $x$ is achieved much faster than
in the regime of the fixed rapidity. It is  seen also that the behavior 
in the vicinity of $x =0$ is determined by the value of intercept of 
secondary trajectory connected with valence quark,
common for the beam hadron and observed particle.

It is quite evident that the applicability region of the above expressions  
is severely restricted  by the necessity to fulfill the condition of large 
values of the $s_1$ and $s_2$ invariants in the case of the Regge 
approximation. However, the general conclusions on the character of an 
asymptotic behavior of charmed particle spectra, which can be obtained from 
them, on the one hand,  are  quite definite, and, on the other hand, agree 
well with the parton model predictions.

We wish to stress once more, that from general theoretical consideration one 
should expect equal yields of the charmed particle and anti-particle in the 
central region $(x \approx 0)$ when $s \to \infty$:
\begin{equation} 
  E \frac{d^2 \sigma(\,D\,)}{d^3 p} |_{x \approx 0} \approx
  E \frac{d^2 \sigma(\,\bar D\,)}{d^3 p} |_{x \approx 0} 
\end{equation}
As we will show such behavior is agree also with parton model predictions.

\section {\bf Fragmentation mechanism } 

In our previous publications ~\cite{ls1, Likhoded} we developed the 
phenomenological model model, where the hadronization of charmed $c$-quark
is described by the sum of two mechanisms, namely
\begin{eqnarray}
 \frac{d \sigma_H}{dx} =  \frac{d \sigma^{F}_H}{dx} 
 + \frac{d \sigma^{R}_D}{dx}, \label{sig1}
\end{eqnarray}
where the first term corresponds to the charmed quark fragmentation,
 while the second term takes into account the charmed $c$-quark 
interaction with
valence quarks from initial hadrons ({\bf recombination}).

In the  fragmentation mechanism the inclusive cross section for
the charmed hadrons ($D$-mesons) production has the form as follows:
\begin{eqnarray}
E_H \frac{d^3 \sigma^F}{d^3 p_H} \; = \; \int 
E_c \frac{d^3 \sigma(h_1 h_2 \to c X)}{d^3 p_c} \; D(z)
\, \delta(\vec p_H \, - \, z \vec p_c) \, d^3 p_c. \label{f2}
\end{eqnarray}
where $z = |\vec p_H| / |\vec p_c|$ is the fraction of the $c$-quark
momentum carried away by the charmed hadron $H$.

The parameterization of the fragmentation function $D(z)$ 
(for example, in the form from \cite{klp} or~\cite{peters}) can be found 
by fitting to the data from reaction $e^+ e^- \to D(c \bar q) \,X$. 
However, in hadronic collisions the situation is more complicated.
Indeed, the use of the fragmentation function is justified at asymptotically 
large values of the invariant mass of $c \bar c$ pair or high~$p_{\top}$. 
However, in hadronic production of the charmed particles 
the main contribution into inclusive charm production
cross section results from $c$-quarks with low values of the invariant mass of
$c \bar c$ pair ($M_{c \bar c} \; \ge 2 m_c$)~\cite{1}. These quarks dominate 
in the region small~$x$. At the same time there is large
amount of partons from initial hadrons in the same (central) region
of~$x$.
Therefore, the $c$-quark in combination with one of such a parton can easily
to produce a charmed hadron. Such a process occurs practically without any
loss of $c$-quark momentum (i.e. $\vec p_H \approx \vec p_c$). Therefore, in
the small $x$ region one should expect the coincidence of the spectra of 
charmed hadrons and $c$-quarks. Whereas at high $x$ region one may use the 
conventional fragmentation mechanism.

Following these arguments we have proposed the modified form of the
fragmentation function~\cite{Likhoded}
\begin{eqnarray}
 D^{MF}(z, M_{c \bar c}) \sim z^{-\alpha(M_{c \bar c})} (1-z), \label{a2}
\end{eqnarray}
with two additional conditions on $\alpha(M_{c \bar c})$:
\begin{eqnarray}
 \begin{array}{l c l l c  l} 
 \alpha(M_{c \bar c}) & \to & -\infty &  D(z) \to \delta(1-z) & {\rm at} &
 M_{c  \bar c} \, \to \, 2m_c, \label{a3} \\ 
 \alpha(M_{c \bar c}) & \to & \alpha_c \approx -2.2 & 
D(z) \to  z^{-\alpha_c} (1-z) & {\rm at} &
 M_{c  \bar c} \, \approx \, M_0. \label{dt3}
 \end{array} 
\end{eqnarray}
The explicit form of $\alpha(M_{c \bar c})$ equal
\begin{eqnarray*}
 \alpha(M_{c \bar c}) = 
 \frac{ 1 - 3 \mu(M_{c \bar c})}{1 - \mu(M_{c \bar c})},
\end{eqnarray*}
where
\begin{eqnarray*}
 \mu(M_{c \bar c}) \, = \, \left ( \frac{ \ln ( \frac {M_{c \bar c}}{2m_c}
 q_0)} {\ln q_0} \right )^{0.464}, \quad  q_0 \approx  0.12
\end{eqnarray*}

Note, that the usage of the fragmentation function assumes the
absence of the interaction of the produced heavy $c$-quark with the 
remnants of the initial hadrons. Therefore, it should be no difference between
the spectra of charmed and anti-charmed hadrons. Moreover, any modification of
the fragmentation mechanism can not reproduce the production
asymmetry (the leading particle effect).

\section {\bf Recombination mechanism } 

The fragmentation mechanism can be apply for the 
production of the $c \bar c$ pair in the color--singlet  state or for high 
$p_T$ production of the open charm. On the other hand, for the case of the 
hadronic production of color $c \bar c$ pair with
small $p_T$ one should takes into account the possibility of charmed
$c$ and $\bar c$ quarks interaction with the initial hadron remnants. 
Therefore, due to the different valence quarks in the initial hadrons 
one may expect the different inclusive spectra of the final charmed hadrons.

In the parton model framework, a heavy $c$--quark should interact with a 
high probability with its nearest neighbor in the rapidity space able to
form a color-singlet state with it. In some cases, the heavy
anti-quark (quark)  may find itself close (in rapidity space) to a valence
light quark (diquark) from the initial hadron. This would result in the 
formation of a fast heavy meson (baryon) in the fragmentation region of the 
initial hadron. 

We use the model \cite{ls1, Likhoded} to describe the production 
asymmetry for charmed hadrons. 
In this model the interaction of the charmed quarks with valence
quarks from the initial hadrons describes with the help of the recombination 
function \cite{ls1, Likhoded}. 
The recombination of the valence $q_V$ and 
$\bar c$ quarks into $D$-meson is described by the function of 
$R_M(x_V, z; x)$:
\begin{eqnarray}
 R_M(x_q, z; x) =  \frac{\Gamma(2 - \alpha_q-\alpha_c)}
 {\Gamma(1-\alpha_c) \Gamma(1-\alpha_q)}
 \xi_q^{(1-\alpha_q)} \xi_c^{(1-\alpha_c)} \; \delta(1 - \xi_q - \xi_c), 
\label{rec1} 
\end{eqnarray}
where $\xi_q = x_q / x$ and $\xi_c = z / x$, while $x_q$, $z$, and $x$ are the
fractions of the initial-hadron c.m. momentum that are carried away by the
valence $q$-quark, charmed $c$-quark, and the $D(\bar c q)$-meson, 
respectively.
The corresponding recombination of three quarks into baryon can be described
by means of the similar recombination function:
\begin{eqnarray} \label{rec2}
 R_B(x_1, x_2, z; x) &=& 
 \frac{\Gamma(3 - \alpha_1-\alpha_2-\alpha_c)} 
 {\Gamma(1-\alpha_1) \Gamma(1-\alpha_2)\Gamma(1-\alpha_c)} \\
 &\times& \xi_1^{(1-\alpha_1)}  \xi_2^{(1-\alpha_2)} \xi_c^{(1-\alpha_c)} \; 
\delta(1 - \xi_1 - \xi_2 - \xi_c). \nonumber 
\end{eqnarray}
These functions take into account the momentum  conservation and the 
proximity of partons in the rapidity space. Actually,
the recombination function is the modulus squared of the heavy meson (baryon)  
wave function in momentum space, being considered in the infinite momentum 
frame in the valence quark approximation. 

With the help of the function $R(x_V, z; x)$ describing the recombination of
the quarks $q_V$ and $\bar c$ into a meson, we represent the corresponding
contribution to the inclusive spectrum of $D$ meson as 
follows~\cite{ls1, Likhoded}:
\begin{eqnarray}
x^{\ast} \frac{d \sigma^{R} }{dx} = R_0 \int x_V z^{\ast} 
\frac{d^2 \sigma}{dx_V dz} \; 
R(x_V, z ; x) \frac{dx_V}{x_V} \frac{dz}{z}, 
\label{sigg1} 
\end{eqnarray}
where $x^{\ast} = 2 E / \sqrt{s}$ and $x = 2 p_l / \sqrt{s}$ (here, $E$ and
$p_l$ are the energy and longitudinal momentum of the $D$-meson in
the c.m.s. of the initial hadrons); $x_V$  and $z$ are the momentum fractions
carried away by the valence quark and heavy anti-quark, respectively, 
and $x_V z^{\ast} \frac{d^2 \sigma}{d x_V \; dz}$ is the double-differential
cross section for the simultaneous production of the quarks $q_V$ and
$\bar c$ in a hadronic collision. 
The equation describing the production of charmed baryon has an analogous form.

The parameter $R_0$ is the constant term of the model, that 
determines the relative contribution of recombination. We fit the data on
charmed hadrons production in $\pi^- N$ collisions and found that 
 $R_0 \approx 0.8$~\cite{Likhoded}.

Note, that  using of the recombination with the valence quarks provides a well
description of the leading particle effect. At the same time its contribution 
in the total inclusive cross section production of the charmed particles is 
sufficiently small~($\sim 10\%$).  This mechanism
dominates in the high $x$ region.

Note, that this model provides more or less successful description of the 
charmed $D$-meson production in $\pi^- \, N$-interactions
(see Fig.~2,~3 and \cite{Likhoded} for details).

\section {\bf Charm production in $\Sigma$--beam }

In this section we consider the charmed hadron production in high-energy beam
of $\Sigma^-$-hyperon. First of all we expect the various behavior of the 
distributions of the valence  $d$- and $s$-quarks.
Indeed, as a first approximation, the distribution of valence quark in the 
baryon
$B(q_1 q_2 q_3)$ can be presented as follows~\cite{ls1, Likhoded}:
\begin{equation}
V^B_{q_1}(x) \propto  x^{-\alpha_1} (1-x)^{\gamma_b - \alpha_2 - \alpha_3},
 \label{vq}
\end{equation}
where $\alpha_i$ is the intercept of the leading Regge-trajectory for
$q_i$-quark, while $\gamma_B \simeq 4$. Note, that due to violation flavor
$SU(N)$-symmetry, we have different intercepts  for
$d(u)$- and $s$-quarks~\cite{collin,klp}:
\begin{eqnarray}
\alpha_u = \alpha_d = \frac{1}{2}, \quad \alpha_s \approx 0, \quad
  \alpha_c \approx -2.2. \label{rec22}
\end{eqnarray}
As a result, the $x$-dependence of the valence $d$ and $s$-quark in the
$\Sigma^-(sdd)$-hyperon has the form as follows~\cite{Kiselev:1986at}:
\begin{equation}
V^{\Sigma}_d \sim \frac{1}{\sqrt{x}} (1-x)^{3.5}, 
\quad
V^{\Sigma}_s \sim (1-x)^{3}    \label{vds}
\end{equation}
It is seen from the~(\ref{vds}) that the valence $s$-quark in the 
$\Sigma^-$-hyperon has harder $x$-distribution than that for 
$d$-quark. Note, that gluon distribution in $\Sigma$-hyperon has the same
form as in usual nucleon. As a result, we expect
more harder $x$-dependence of $s$-quark distribution in $\Sigma$-hyperon
then $u$-quark distribution in nucleon. Thus, one may expect slightly
harder spectrum of $D_s$-mesons, produced in $\Sigma$-beam, then spectrum of
$D^0$-mesons, produced in $nucleon$-beam. On the other hand, $d$-quark in 
$\Sigma$-hyperon should be slightly softer then that in nucleon.
 As a result we should
observe the different $x_F$-dependence of spectra of charmed hadrons with
$d$- or  $s$-quarks, namely, $D^-(\bar c d)$ and $D_s^-(\bar c s)$,
$\Xi_c^0(cds)$ and $\Sigma_c^-(cdd)$, etc.

We use LO formulas for calculations of the cross sections for quark-antiquark 
and gluon-gluon annihilation into $c \, \bar c$-pair. We set $m_c=1.25$~GeV 
and  the value of strong coupling constant equals~0.3.
Then for the cross section value of $\Sigma^- \, p$ interaction
at $P_{LAB} = 600$~GeV one has:
\begin{equation}
 \sigma(\Sigma^- \, p \, \to \, c \, \bar c \, X) \simeq 8\,\,\mu{b}
\end{equation}
In our calculations we do not pretend to reproduce the absolute value of
this cross section (see~\cite{1}, for detail consideration of this problem).
We concentrate on the description of $x_F$-distribution of charmed mesons and
baryons. 

Recently, we have done the calculations of asymmetry of $x$-spectra for
$D$, $D_s$ and $\Lambda_c$ hadrons produced in $\Sigma^-$ beam with the energy
of 340~GeV. The WA89~Collaboration have compared these predictions with
their experimental data~\cite{Adamovich:1999mu}. This comparison 
is presented in Fig.4.
As is seen from the picture the asymmetry in $D$-meson production in 
$\Sigma$-beam is different from that in $\pi^-$-beam and  there is more 
pronounced asymmetry for $D_s$-meson production.

Below we present the predictions for charmed hadrons production in 
$\Sigma^-$-beam at 600~GeV energy. 
The corresponding distributions (integrated over $p_\top$) are 
presented in Fig.5. We may see from these figures, that the considered
charmed quark interaction in the final state (recombination)
leads, indeed, to noticeable differences in $x_F$-spectra. These differences
can be explicitly seen in Fig.6, where we present the corresponding asymmetry
$A$ (see~(\ref{asym}) for definition).
The most non-trivial prediction of the proposed model is presented in two
lower plots in Fig.6,
where we present the ratio of the inclusive spectra of $D^{*-}_s(\bar c s)$
and $D^{*-}(\bar c d)$ mesons as well as $\Xi_c^0$ and $\Sigma_c^0$ baryons.
Indeed, due to the difference of the valence $d$ and $s$ quarks in the 
$\Sigma^-$-beam (see~(\ref{vds})) we expect the additional asymmetry in the
leading charmed particles production.

\vspace*{0.5truecm}
\section {\bf CONCLUSION } 

In the present article we wish to stress once more, that the source of the 
observed
asymmetry in charmed hadron production is the interaction of produced charmed
quarks with valence quarks from initial hadrons. Note,
the model under consideration provides also the additional method
to measure the valence quark distribution functions of $\Sigma$-baryons.

\vspace{0.8cm}

\noindent {\bf ACKNOWLEDGMENTS }

\noindent The authors thank E.~Chudakov, A.~Kushnirenko, O.~Piskunova, and 
J.S.~Russ for the fruitful discussions.

This work was supported in part by Russian Foundation for Basic Research,
projects no.~~99-02-16558.

\vspace*{1cm}
\noindent {\bf P.S.} When this article was finished we received the 
article~\cite{Aitala:2000rd} with the recent results of 
E791~Collaboration. They present the inclusive $x$ and $p_{\top}$ 
distributions of charmed hadrons, produced in $\pi^-$-beam.
In particular, they observed a noticeable asymmetry in $\Lambda_c^+$ 
and $\overline{\Lambda_c^-}$ baryons production in the forward region 
(about 13\%).
Note, that our model can not explain this result. Due to the equal amounts of 
valence ${\bf \bar u}$ and ${\bf d}$ quarks in the $\pi^-$-beam we expect 
an equal yields of the $ \Lambda_c^+(c u {\bf d})$ and
$\overline{\Lambda_c^-(c {\bf u} d)}$ baryons in the $\pi^-$-beam.
Moreover, the most theoretical models~\cite{models} predicts also zero
asymmetry for these particle production.

\newpage
\begin{center}
\begin{figure}[t]
\begin{center}
\epsfig{file=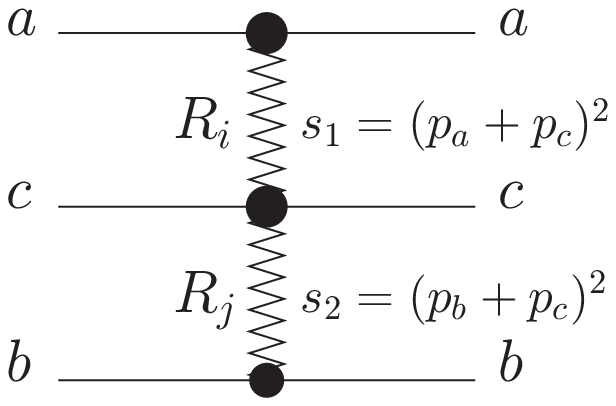,width=8.cm,clip} 
\end{center}
\ccaption{}{
 Six particle $M_{3\to3}$ amplitude.
 }  \label{fig1}
\end{figure}
\end{center}

\begin{figure}[t]
\epsfig{file=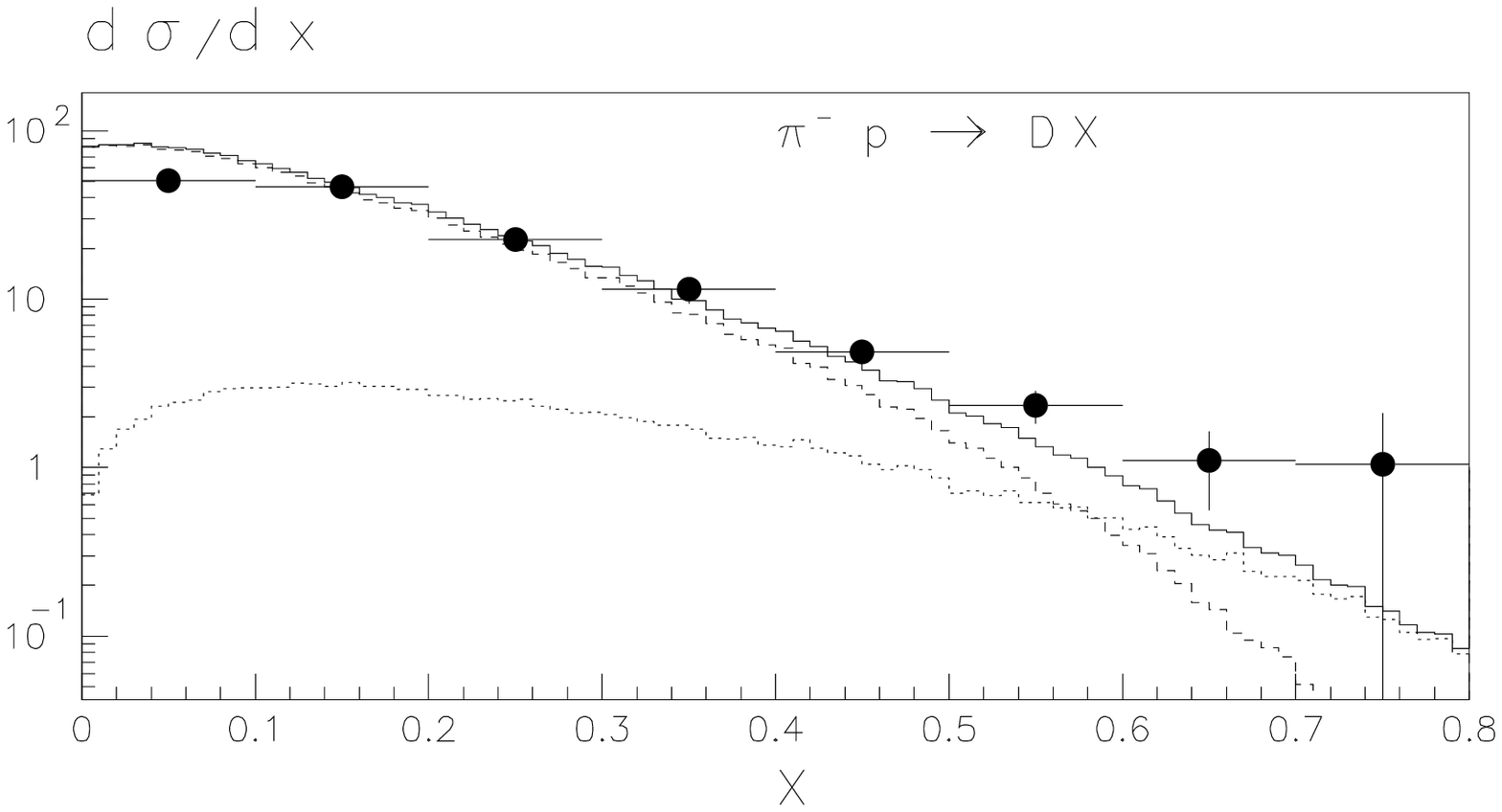,width=14cm,clip} 
\ccaption{}{
Differential distributions $\frac{d \sigma}{dx}$ for
the energy of $E_{\pi} = 250$~GeV~\cite{expall}. 
The dotted (dashed) histogram corresponds to the recombination (fragmentation)
contribution. The solid histogram represents their sum. The cross sections are
presented in $\mu$b (see~\cite{Likhoded}
for details).  }  

\epsfig{file=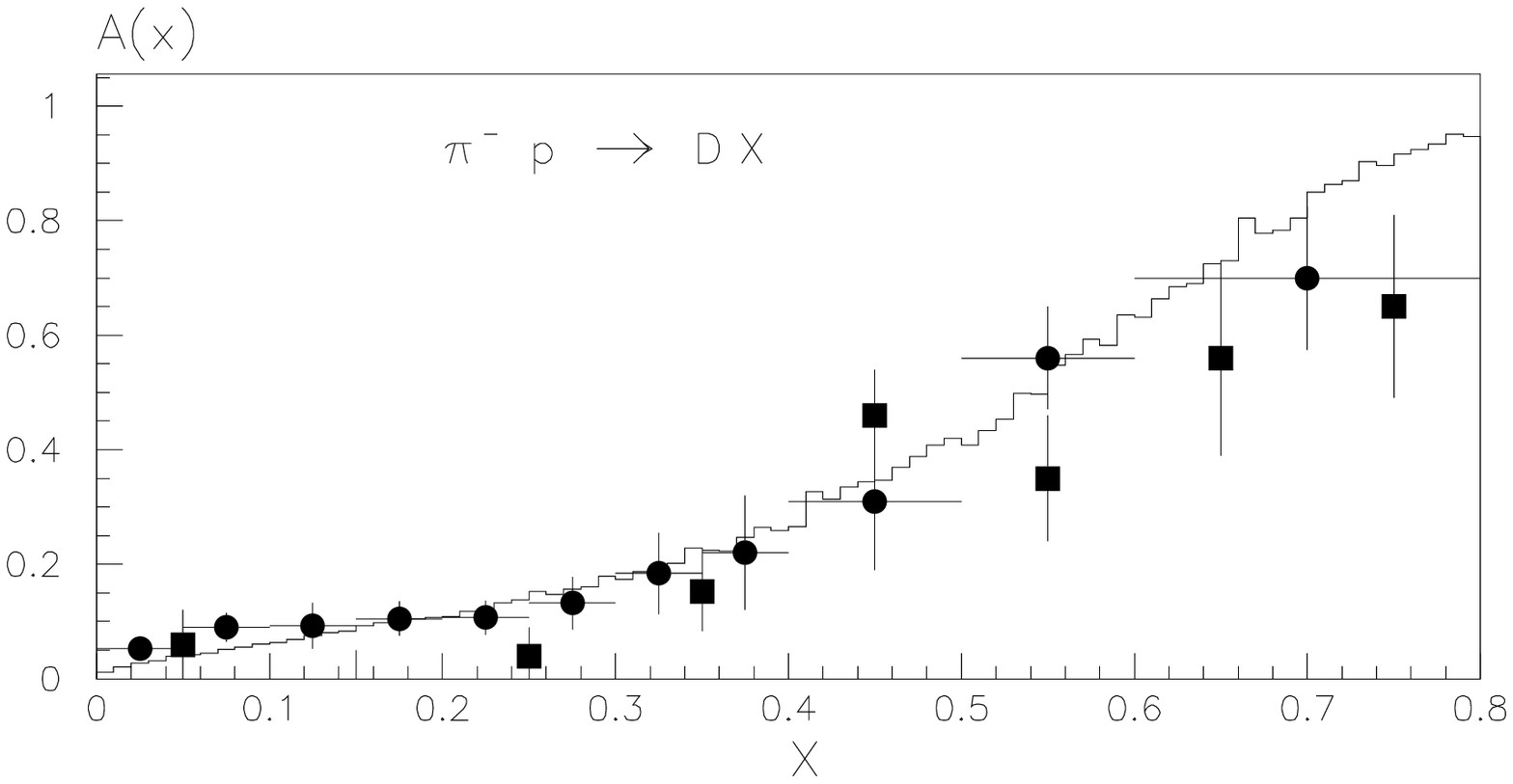,width=14cm,clip} 
\ccaption{}{ The description of the asymmetry $A(x)$ 
in $\pi^- p $ collisions (see~\cite{Likhoded} for details) } 
\end{figure}

\newpage

\begin{figure}[t] 
\epsfig{file=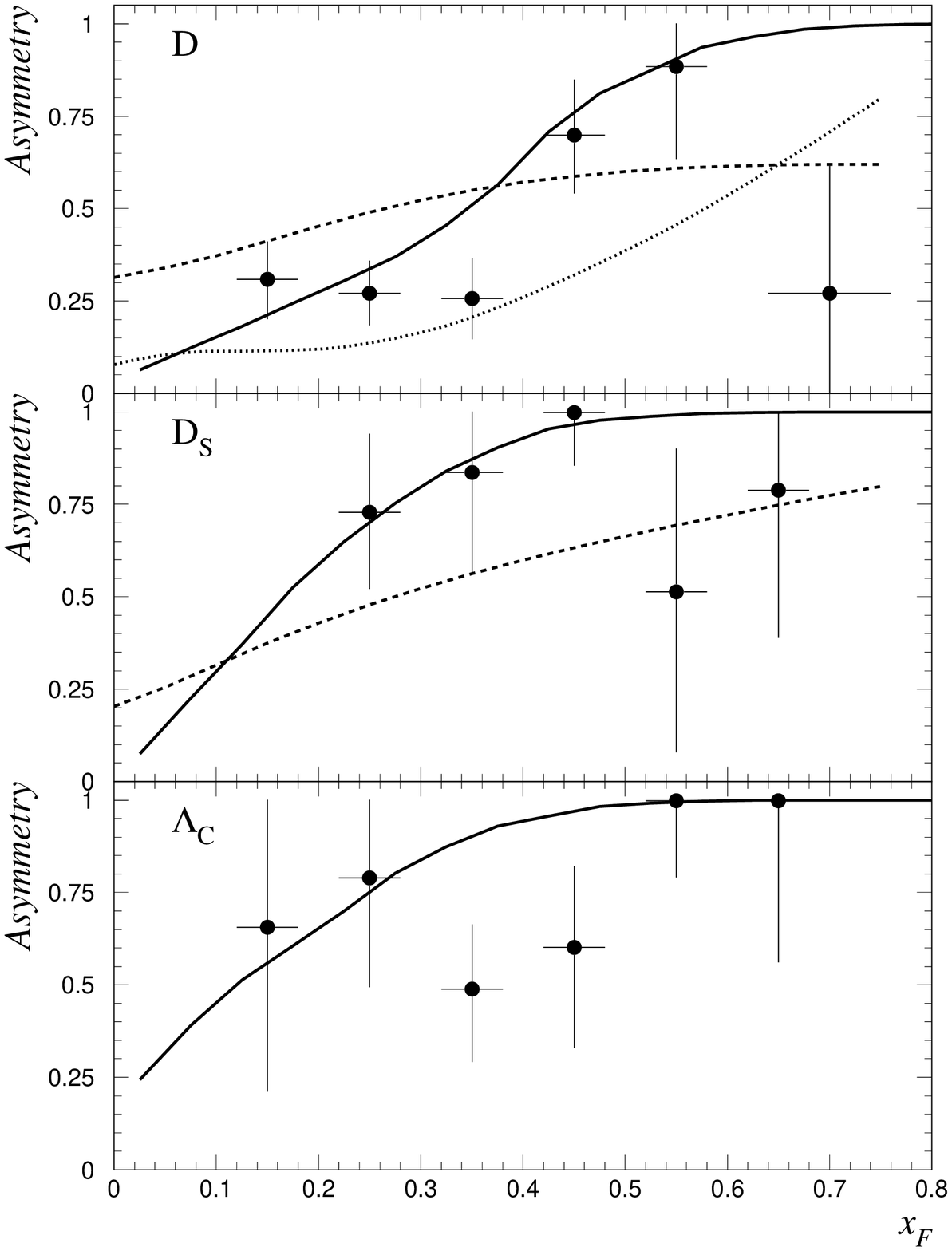,width=12cm} 
\ccaption{}{ The asymmetry in the  charmed hadrons
produced in $\Sigma^- \, p$ interactions at 
$P_{LAB} = 340$~GeV~\cite{Adamovich:1999mu}.
 }
\end{figure}

\newpage

\begin{figure}[t]
\epsfig{file=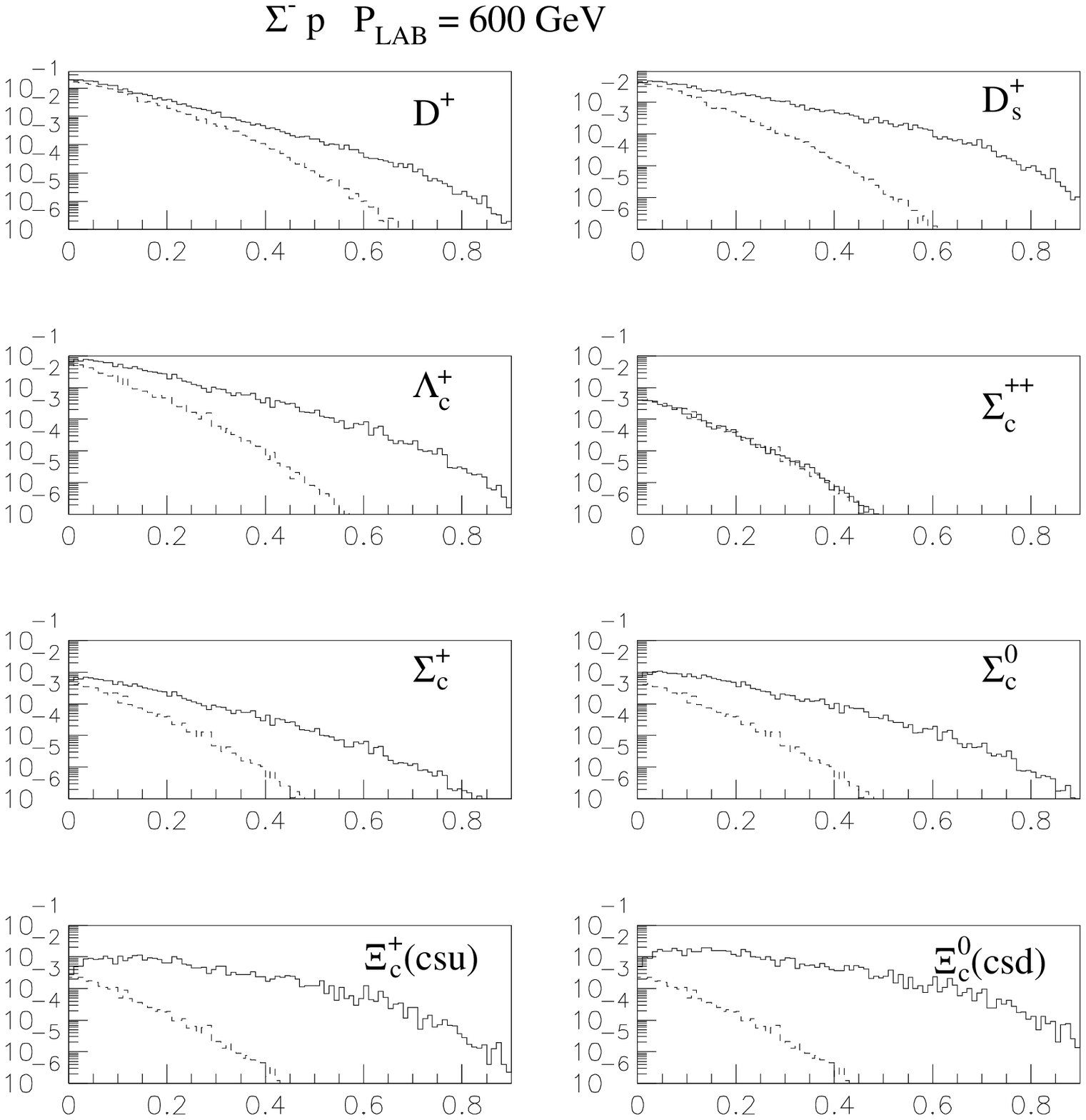,width=16cm} 
\ccaption{}{$x$--distributions of charmed mesons ($D^{* \pm}$ and 
$D_s^{* \pm}$) and baryons
produced in $\Sigma^- \, p$ interactions at $P_{LAB} = 600$~GeV. 
The solid (dashed) curves correspond to leading (non-leading) charmed hadron
production. } 
\end{figure}

\newpage

\begin{figure}[t]
\epsfig{file=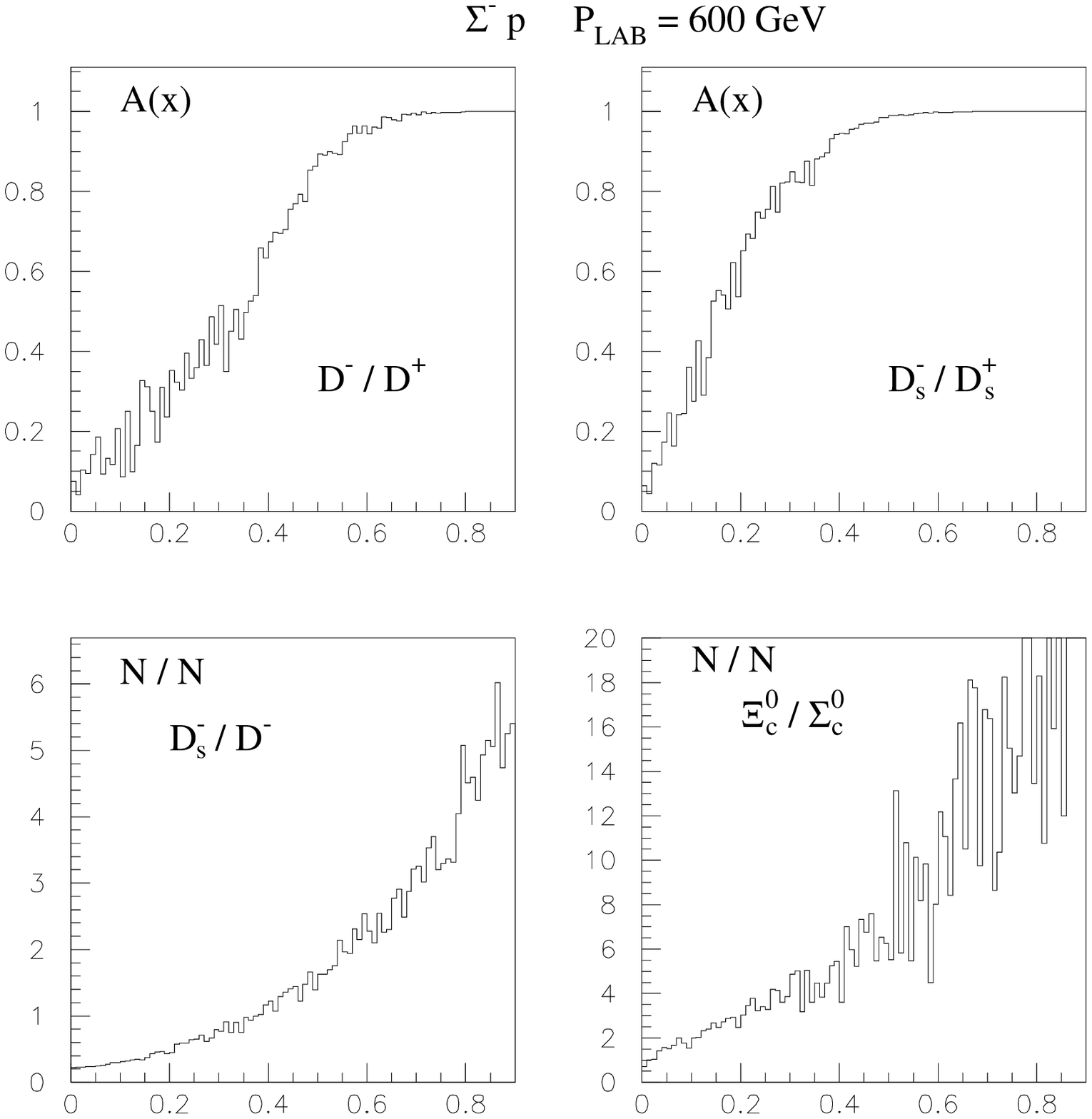,width=16cm} 
\ccaption{}{ Production asymmetry $A(x)$~from~(\ref{asym}) 
for $D^{* \pm}$ and $D_s^{* \pm}$ mesons (two upper histograms),
 produced in $\Sigma^- \, p$ interactions at $P_{LAB} = 600$~GeV.
Two lower plots present the ratio of the spectra
of charmed hadrons with and without strange quarks
(i.e. the $D^{* -}_s / D^{* -}$ and 
$\Xi^{0}_c(cds) / \Sigma^0_c(cdd)$ ratios).   } 
\end{figure}

\end{document}